\documentstyle[aps]{revtex} 
\begin{document}
\draft
\title{Mass and charge fluctuations and black hole entropy} 
\author{Ashok Chatterjee\footnote{email: ashok@theory.saha.ernet.in} 
and Parthasarathi Majumdar\footnote{email:
partha@theory.saha.ernet.in}} 
\address{Theory Group, Saha Institute
of Nuclear Physics, Kolkata 700 064, India.} 
\maketitle
\begin{abstract} 
The effects of thermal fluctuations of the mass
(horizon area) and electric charge, on the entropy of non-rotating
charged {\it macroscopic} black holes, are analyzed using a grand
canonical ensemble. Restricting to Gaussian fluctuations around
equilibrium, and assuming a power law type of relation between the
black hole mass, charge and horizon area, characterized by two real
positive indices, the grand canonical entropy is shown to acquire a
logarithmic correction with a positive coefficient proportional to
the sum of the indices. However, the root mean squared fluctuations 
of
mass and charge relative to the mean values of these quantities turn
out to be independent of the details of the assumed mass-area 
relation. We also
comment on possible cancellation between log (area) corrections
arising due to {\it fixed area} quantum spacetime fluctuations and
that due to thermal fluctuations of the area and other quantities.
\end{abstract} 
\vglue .2in

\section{Introduction}

The notion of entropy of a spacetime such as a black hole presumes
the existence of microstates whose degeneracy $g$ is measured by that
entropy. It is conceptually difficult to think of microstates of a
{\it classical} stationary spacetime. Yet even vacuum solutions of
Einstein's equation like the Schwarzschild and the Kerr black holes
are supposed to possess an entropy equal to a quarter of their
horizon area (in Planckian units) \cite{bek}, \cite{haw}. The only
way in which microstates might arise in these cases is within a
quantum mechanical description of such spacetimes themselves (and not
merely of the matter fields propagating in them) which allow for
dynamical fluctuations. In other words, the entropy of black holes
must be taken to originate from quantum degeneracy of spacetime
geometry, and consequently, can only be properly analyzed within a
quantum theory of gravitation \cite{bek}.

Of all attempts to construct a `quantum general relativity', one that
perhaps comes closest to dealing directly with fluctuations of
spacetime geometry (within a canonical framework) is Loop Quantum
Gravity (LQG) \cite{al}. Yet even in this approach, the only 
situation 
that can be
reliably analyzed is one in which the black hole is {\it isolated},
i.e., has a horizon with a {\it fixed} (macroscopic) area ${\cal A} >>
1$ (in units of Planck area). From a statistical
thermodynamics standpoint, this corresponds to employing a {\it
microcanonical} ensemble to compute the entropy \cite{aa1}. Microstates
leading to the microcanonical entropy ${\tilde S}_{MC} \equiv \log g$ 
are identified with
boundary states of a three dimensional Chern-Simons theory `living' on
the horizon \cite{aa1}. The local Lorentz invariance is gauge-fixed to 
a local $SU(2)$ on an appropriate spatial
slice. Thus one may take this $SU(2)$ to be the gauge group of the
Chern Simons theory. However, boundary conditions associated with the
`isolatedness' of the horizon have been argued \cite{aa1} to further
gauge-fix the $SU(2)$ to a residual $U(1)$. In either case, the
degeneracy of the boundary states of the Chern Simons theory yields
(for macroscopically large horizon areas) a term logarithmic in the 
area together with an infinite series
in inverse powers of the area \cite{km1}, \cite{km2}, over and above 
the Bekenstein-Hawking entropy $S_{BH}=A/4$ (in Planckian units),
\begin{eqnarray}
{\tilde S}_{MC} ~&=&~ S_{BH} -\frac32 \log S_{BH} + const.+ 
O(S_{BH}^{-1})~~\mbox{for $SU(2)$} \nonumber \\ 
       &=&~ S_{BH} -\frac12 \log S_{BH} + const.+ O(S_{BH}^{-1}) 
~~\mbox{for $U(1)$} 
\label{smc}
\end{eqnarray}
The negative sign of the dominant logarithmic correction in both
cases is due to the fact that invariance of the horizon states under
the gauge group effectively reduces the degeneracy as compared to the
leading order term (area law) where the invariance is not taken ito
account. Notice that each term of the infinite series in eq.
(\ref{smc}) is {\it finite and unambiguously calculable}.

The problem with an isolated black hole is that it neither radiates nor
absorbs anything. This is unsatisfactory from a physical standpoint
since classically a black hole absorbs everything, and with quantum
matter around, it radiates all kinds of quanta in a Planckian
distribution \cite{haw} at the Hawking temperature given by the
surface gravity on the horizon. Thus, the appropriate equilibrium
ensemble to probe black hole thermodyamics ought to be the canonical
(Gibbs) ensemble (generalized to the grand canonical ensemble when
fluctuations of charge or angular momentum are included).  
Unfortunately, however, for asymptotically flat black hole spacetimes,
the equilibrium is unstable, because of a superexponential growth of
the density of states leading to a diverging canonical partition
function. So, depending on whether the ambient temperature is larger or
smaller than the Hawking temperature, the black hole either accretes or
radiates without limit.

The situation is more favourable for asymptotically anti-de Sitter
black holes which, for a certain range of parameters, can indeed be in
stable thermal equilibrium \cite{hawp}. For such spacetimes the area
(energy) fluctuations within a canonical ensemble lead, within a 
saddle-point approximation, to an additional  
correction term in the canonical entropy, over and above the 
infinite series of corrections to the area law found for the 
microcanonical entropy \cite{cm1},
\begin{eqnarray}
\delta_{th}S_C~=~ \frac12 \log S_{BH} ~-~ \frac12
\log(n-1)~+~const~,\label{ther}
\end{eqnarray}
where $n$ is the index appearing in the assumed power law relation 
between area of the horizon and energy. This additional correction has a 
positive sign, as expected from the fact that thermal fluctuations 
tend to disorder the system and therefore  
always increase the canonical entropy. The area-dependent part is 
universal in that it is independent of the index characterizing the 
asuumed mass-area power-law relation. We may remark that 
eq. (\ref{ther}) is valid in the limit of large horizon areas and for 
fixed (i.e. thermally non-fluctuating) charge or angular momentum of 
the black hole. For asymptotically flat black holes, $n=1/2$, implying 
that the entropy as well as the free energy acquires an imaginary part. 
This indicates a breakdown of the saddle-point approximation used to 
compute the canonical entropy from the canonical partition function. 
It is not unlikely that this breakdown is indicative of the thermal 
instability ensuing in the canonical ensemble for such spacetimes, as 
mentioned earlier. On the other hand, for {\it all} anti-de Sitter 
black 
holes, $n=3/2$, and therefore the calculation above is reliable. 

One immediate fallout of this additional correction is that it begins
to compete with the microcanonical corrections to the area law. The 
net canonical entropy now takes the form 
\begin{eqnarray}  
S_C ~ &=& ~ {\tilde S}_{MC}~+~\delta_{th}~S_C \nonumber \\
&=& ~ S_{BH} ~-~ \log S_{BH} ~-~ \frac12 \log(n-1) ~+~ const.
~+~\dots
~.\label{sct}
\end{eqnarray}
for the $SU(2)$ case, while for the $U(1)$ case, the two 
leading logarithmic corrections to the area law simply cancel each 
other out \cite{cm2}. At this point the cancellation is not much more 
than a point of curiousity, since there is still an infinite series of 
other (power-law) corrections hidden in the $\dots$ in 
eq. (\ref{sct}). However, if such cancellations are proven to work 
for all orders in powers of inverse area, one may ponder about 
symmetries, non-renormalization theorems etc. 

Recall that so far the discussion of the canonical entropy has assumed
fixed charges/angular momenta of the black holes. In this paper we
consider the effect of including thermal fluctuations of the electric
charge for non-rotating spherically symmetric black hole spacetimes. In
other words, we consider the generalization of the previous canonical
ensemble formulation to a grand canonical ensemble framework where the
electrostatic potential of the black hole plays the role of a chemical
potential. Thus, the heat bath which the black hole interacts with is
now electrically charged. The use of a saddle-point approximation in
this case involves important technical changes which are incoporated in
our treatment. The object here is to determine whether the universality
found in the canonical treatment survives the generalization. Once
again, we make no more assumption than is absolutely essential. In
particular, we adhere to the area spectrum strictly as derived in the
LQG formulation \cite{al}, specialized to the case of large macroscopic
areas where the spectrum becomes equally spaced.\footnote{Only in one
other approach, involving a proposal for coherent states \cite{thie}
within LQG, is there a proper {\it derivation} of an equally-spaced
area spectrum \cite{ad} which does not involve ad hoc assumptions
regarding the black hole mass spectrum. Such ill-founded assumptions
often lead to claims regarding corrections to the area law \cite{hod}
which we find spurious.}

In this context, we may note that two recent papers \cite{gm},
\cite{med} have explored the effect of thermal energy and charge
fluctuations for a grand canonical ensemble of AdS Reissner-Nordstrom
black holes. While our result for the leading thermal fluctuation
correction to the grand canonical entropy for this particular black
hole matches with the results quoted in \cite{gm}, the implied
assumption in those papers regarding the independence of energy and
charge fluctuations, as also about their relative size, are not
tenable even for large black holes, as we shall show in what follows.
Consequently, the conjectured complete cancellation (including
angular momentum fluctuations) between the microcanonical and 
thermal logarithmic corrections to the area law, as proposed in
\cite{med}, appears to be without basis.
 
The paper is organized as follows: in section 2 we present a grand
canonical ensemble formulation of small fluctuations for general
equilibrium statistical mechanical systems. This is then used in
section 3 to compute the leading logarithmic corrections to the grand
canonical entropy, via a Poisson resummation procedure followed by a
saddle point approximation to perform the integrals in the grand
canonical partition function. In section 4. we study the relative
sizes of fluctuations in more detail to show that they indeed ensure
that the approximation of small Gaussian fluctuations is valid for
large area black holes, when the charge is also allowed to grow with
area in a certain way. However that is shown not to be true for black
holes of vanishingly small charge considered in \cite{gm}. We discuss
our results and conclude in section 5.

\section{Mass and charge fluctuations in the grand canonical ensemble}

For a thermodynamic system with Hamiltonian ${\hat H}$ and charge 
operator ${\hat Q}$ in equilibrium with a heat and charge reservoir at 
temperature $T$ and electrostatic potential $\Phi$, the grand partition 
function is 
\begin{eqnarray}
Z_G~~=~~Tr \exp - \beta ({\hat H}~-~\Phi {\hat Q})~ \label{gpf}
\end{eqnarray}
When the spectra of ${\hat H}$ and ${\hat Q}$ are both continuous with 
eigenvalues $E$ and $Q$ respectively, the partition sum in eq.  
(\ref{gpf}) can be conveniently rewritten as an integral
\begin{eqnarray}
Z_G~~=~~\int dE ~dQ ~\rho(E,Q)~\exp -\beta(E-\Phi~ Q)~, \label{gpfint}
\end{eqnarray}
where the density of states $\rho(E,Q)$ may be taken to be a smooth 
function and $\beta \equiv T^{-1}$. 

Defining the microcanonical entropy $S_{MC}(E,Q) \equiv \log
\rho(E,Q)$,\footnote{Observe that $S_{MC}$ defined as above is
precisely what we called ${\tilde S}_{MC}$ in \cite{cm1} where we
defined $S_{MC} \equiv \log g(E)$ with $g(E)$ being the degeneracy of
states with energy $E$. This distinction between the two definitions 
of
microcanonical entropy is relevant for logarithmic corrections to the
area law \cite{cm1}.} we get
\begin{eqnarray}
Z_G~~=~~\int dE ~dQ ~\exp ~ \left\{ S_{MC} -\beta~(E-\Phi ~Q) \right\} 
~, \label{gpf2}
\end{eqnarray}
Assuming that the integral is dominated by the saddle point of the 
exponent at $E=M,~Q=Q_0$, we expand the exponential in the integrand 
upto terms 
quadratic in the fluctuations $\delta E \equiv E-M~,~\delta Q \equiv
Q-Q_0$
\begin{eqnarray}
S_{MC}(E,Q)-\beta (E-\Phi ~Q) &=& S_{MC}(M,Q_0)-\beta(M-\Phi~ Q_0)  
\nonumber \\
&+& {1 \over 2}  \left[ ~\left( S_{MC,EE} \right) _{M,Q_0}~ \delta 
E^2 
+ \left( S_{MC,QQ} \right) _{M,Q_0}~ \delta Q^2 + 2~ 
\left( S_{MC,EQ} \right)_{M,Q_0}~\delta E \delta Q ~\right] ~, 
\label{expn}
\end{eqnarray}
where, $S_{MC,EE} \equiv \partial^2 S_{MC} /\partial E^2$, etc. In 
writing eq. (\ref{expn}) above we have made use of the saddle point 
conditions
\begin{eqnarray}
\left( S_{MC},E \right)_{M,Q_0}~ &=&~ \beta ~=~T^{-1} \nonumber \\
\left( S_{MC},Q \right)_{M,Q_0}~ &=&~ -\beta~\Phi = -\Phi/T 
~\label{sad}
\end{eqnarray}
which are the usual relations for the temperature  and potential in the 
microcanonical ensemble. 

The presence of cross terms involving $\delta E$ and $\delta Q$ in 
the exponent in eq.(\ref{gpf2}) (using (\ref{expn}) implies that the 
energy
and charge fluctuations cannot, in general, be treated independently. 
Evaluating the Gaussian integrals over the fluctuations yields,
\begin{eqnarray}
Z_G~=~{2\pi \over \sqrt{det~\Omega}}~\exp~ \left\{ S_{MC}(M,Q_0) - 
\beta M + \beta \Phi Q_0 \right\} ~, \label{gpf3}
\end{eqnarray}
where, the Hessian matrix
\begin{eqnarray}
\Omega ~=~ \pmatrix{
S_{MC,EE} &  S_{MC,EQ} \cr
S_{MC,EQ} &  S_{MC,QQ} }_{M,Q_0} ~\label{omeg}
\end{eqnarray}
is assumed to be negative definite to ensure stability under small 
fluctuations. The necessary and sufficient conditions for this are 
\begin{eqnarray}
Tr \Omega ~&=& ~ S_{MC,EE} \left|_{M,Q_0}~+~S_{MC,QQ} 
\right|_{M,Q_0}~ < ~ 0 \nonumber \\
\det \Omega ~&=& ~ \left[ ~S_{MC,EE}~S_{MC,QQ}~-~S_{MC,EQ}^2 ~
\right]_{M,Q_0}~>~ 0 ~
\label{posd}
\end{eqnarray}
which necessarily imply 
\begin{eqnarray}
S_{MC,EE}|_{M,Q_0}~<~0~,~ \mbox{and}~  S_{MC,QQ}|_{M,Q_0}~<~0~. 
\label{psd}
\end{eqnarray}
Note that while these conditions together imply the first of the
necessary and sufficient conditions (\ref{posd}) for stability of
$\Omega$, they are not sufficient to guarantee the second one.

Using the microcanonical relations for temperature and potential, we 
may expess the necessary conditions for stability in terms of the heat 
capacity $C_Q \equiv (\partial E/\partial T)_Q$ and the capacitance $C 
\equiv (\partial Q / \partial \Phi)_E$ in the following way
\begin{eqnarray}
C_Q~~ > ~~ 0~~ \mbox{and}~~ C~\Phi~~ < ~~ T~\left( \partial Q \over 
\partial T \right)_E ~. 
\label{capa}
\end{eqnarray}
The more stringent necessary and sufficient conditions can also be 
similarly expressed in terms of $C_Q$ and $C$. 

The grand partition function, evaluated in the saddle point 
approximation, can now be substituted in the standard thermodynamic 
relation in the presence of a chemical (electrostatic) potential 
\begin{eqnarray}
S_G~ = ~ \beta~M~-~\beta~Q~\Phi~+~ \log Z_G ~, \label{sg}
\end{eqnarray}
so as to yield the grand canonical entropy
\begin{eqnarray}
S_G~~=~~ S_{MC}~-~\frac12~ \log ~\det ~\Omega ~+~const. ~ 
\label{gentro}
\end{eqnarray}

It is pertinent at this point to probe the validity of the saddle 
point approximation because it is used throughout this paper. One way 
to do this is to use this approximation to compute the mean-squared 
fluctuations of the energy and charge, 
\begin{eqnarray}
\Delta E^2 ~& \equiv &~ \langle \delta E^2 \rangle ~=~ Z_G^{-1}~\int~ 
dE~dQ~(E-M)^2~e^{S_{MC}- \beta E + \beta \Phi Q } \nonumber \\
\Delta Q^2 ~& \equiv &~ \langle \delta Q^2 \rangle~= ~Z_G^{-1} 
~\int~dE~dQ~(Q-Q_0)^2~e^{S_{MC} - 
\beta E + \beta \Phi Q } \label{fluct}
\end{eqnarray}
and to ensure that these have 
certain desirable properties, e.g., that they are both positive and 
relatively small. Computing appropriate Gaussian integrals, it is 
easy to show that \footnote{For a similar analysis of energy and 
particle number fluctuations in a grand canonical ensemble, see 
\cite{ll}} 
\begin{eqnarray}
\Delta E^2 ~=~ -~ (\Omega^{-1})_{11} ~&=&~ - ~{ 
S_{MC,QQ}(M,Q_0) \over {\det~ \Omega}}  \nonumber \\
\Delta Q^2 ~=~ -~ (\Omega^{-1})_{22} ~&=&~ - ~ { 
S_{MC,EE}(M,Q_0) \over {\det~ \Omega}}  ~. \nonumber \\
\langle \delta E~\delta Q \rangle~ =~ -~( \Omega^{-1})_{12} 
~&=&~  {S_{MC,EQ}(M,Q_0) \over {\det~ \Omega}} ~. 
\label{sadd}
\end{eqnarray}
Using the inequalities (\ref{posd}), (\ref{psd}) both mean-squared
fluctuations are clearly seen to be positive, as they ought to be.
The cross-correlation term, which is generally nonzero, explicitly
demonstrates why it is erroneous to treat these fluctuations as
independent, as already remarked.

\section{Fluctuation contribution to the entropy of large black 
holes}

The canonical black hole entropy has been argued \cite{cm2} to be
determined by the ADM Hamiltonian which describes quantum
fluctuations {\it on the horizon} treated as an inner boundary of
spacetime, essentially because the bulk Hamiltonian obeys the quantum
Hamiltonian constraint. Unfortunately, the spectrum of this boundary
Hamiltonian is yet to be determined in LQG. Thus, as in \cite{cm1},
we assume that energy spectrum is a function of the discrete area
spectrum (well-known in LQG \cite{al}) and a discrete charge
spectrum. The charge spectrum is of course equally spaced in general;  
for large macroscopic black holes the area spectrum is equally spaced
as well.

In a basis in which both the area and charge operators are
simultaneously diagonal, the grand canonical partition function can
be expressed as
\begin{eqnarray}
Z_G~=~ \sum_{m,n}~g(m,n)~\exp - \beta \left[~ E (A_m,Q_n)~-~\Phi~Q_n 
\right] ~, \label{gpf5}
\end{eqnarray}
where, $g(m,n)$ is the degeneracy corresponding to the area 
eigenvalue $A_m$ and charge eigenvalue $Q_n$. 
Using a generalization of the Poisson resummation formula
\begin{eqnarray}
\sum_{m,n}~f(m,n)~=~\sum_{k,l}~\int~dx~dy ~\exp ~\{-i(kx + 
ly)\}~f(x,y) ~\label{poi}
\end{eqnarray}
and assuming that the partition sum is dominated by the large 
eigenvalues $A_m~,Q_n$, it can be expressed as a double integral
\begin{eqnarray}
Z_G~=~\int ~dx~dy~\exp ~-~\beta ~\{ E(A(x),Q(y))~-~\Phi~Q(y) 
\}~g(A(x),Q(y))~. \label{gpf6}
\end{eqnarray}
Note that the transition from the discrete sum to the integral for 
$Z_G$ requires only that the dominant eigenvalues are large compared 
to the fundamental units of discreteness which for the area is the 
Planck area and for the charge is the electronic charge. These 
conditions are of course fulfilled for all astrophysical black holes. 

Changing variables in eq. (\ref{gpf6}) form $x,y$ to $E,Q$
\begin{eqnarray}
Z_G~ &=& ~\int dE~dQ~{\cal J}(E,Q)~g(E,Q)~\exp~\{- \beta (E- \Phi 
Q) \}~ \nonumber \\
&=&~\int dE~dQ~\rho(E,Q)~\exp~\{ - \beta (E- \Phi Q) \} ~, 
\label{gpf7}
\end{eqnarray}
where, the Jacobian ${\cal J}~=~\left|~ E_{,x} ~\right|^{-1} ~ 
\left| ~Q_{,y}~ \right|^{-1} $, and $\rho = {\cal J}(E,Q)~g(E,Q)$ is 
the density of states. Employing the saddle point approximation and 
using eq. (\ref{sg}) one obtains
\begin{eqnarray}
S_G(M,Q_0)~=~{\tilde S}_{MC}(M,Q_0)~-~\frac12~\log \Delta ~+~const. ~ 
\label{sg22}
\end{eqnarray}
where, using $S_{MC} = \log \rho = {\tilde S}_{MC} + \log {\cal J}$ 
we have defined
\begin{eqnarray}
\Delta ~\equiv~ \det \Omega~{\cal J}^2~& = &~\det 
\Omega~(E_{,x})^2~(Q_{,y})^2|_{M,Q_0} \nonumber \\
& = &~ \det \Omega~(A_{,E})^2 ~(A_{,x})^2~ (Q_{,y})^2|_{M,Q_0} ~, 
\label{del}
\end{eqnarray}
and $\Omega$ is defined in eq. (\ref{omeg}). 

Since the microcanonical entropy is known to be only a function of 
the horizon area even for {\it charged} non-rotating black holes 
\cite{aa1,km1}, one can express $\det \Omega$ as
\begin{eqnarray}
\det \Omega ~ & = & ~(S_{MC,A})^2~\left[ (A_{,EE})~(A_{,QQ}) - 
(A_{,EQ})^2 \right] |_{M,Q_0} \nonumber \\
&+& S_{MC,AA}~S_{MC,A}~\left[ (A_{,E})^2~A_{,QQ} + (A_{,Q})^2~A_{,EE} 
- 2~A_{,E}~A_{,Q}~A_{,EQ} \right] |_{M,Q_0} ~\label{omeg2}
\end{eqnarray}
Now, to obtain the leading thermal fluctuation correction to the area 
law (over and above corrections occurring in $S_{MC}$) it is 
adequate 
to retain only the area term in $S_{MC}$ in eq. (\ref{omeg2}), 
yielding
\begin{eqnarray}
\det \Omega ~=~\frac1{16}~ \left[ A_{,EE}~A_{,QQ}~-~(A_{,EQ})^2 
\right]_{M,Q_0}~. \label{deto} 
\end{eqnarray} 
With both the area and charge spectrum being equally spaced (i.e.,
linear in the quantum numbers $x$ and $y$), we therefore obtain the
leading thermal fluctuation correction
\begin{eqnarray}
\Delta ~=~{C \over 16}~\left[ A_{,EE}~A_{,QQ}~-~(A_{,EQ})^2 \right]~ 
\left( A_{,E} \right)^2|_{M,Q_0} \label{delt2}
\end{eqnarray}
where $C$ is a constant related to the area and charge spectra.

To proceed further, it is necessary to specify the functional
dependence of the area of the horizon $A$ on the energy $E$ and the
charge $Q$. As mentioned earlier, this is not yet known within the
LQG framework. We adopt here the alternative of {\it postulating} an
asymptotic power law relationship between these quantities for large 
$A$
\begin{eqnarray}
E~=~a~A^{\alpha}~+~b~Q^2~A^{-\beta}~, \label{powa}
\end{eqnarray}
where, $a~,~b$ are real positive dimensional coefficients dependent 
on the Newton constant and the cosmological constant, and 
$\alpha~,\beta$ are dimensionless real positive exponents. An 
equivalent way of rewriting eq. (\ref{powa}) is
\begin{eqnarray}
E~=~a~A^{\alpha}~\left[ 1~+~b' Q^2~A^{-\beta'} \right] 
~,  \label{powp}
\end{eqnarray}
where $b'$ is similar to $b$ above and likewise 
for $\beta'$. Both forms of power-law relation subsume {\it 
all} four dimensional charged non-rotating black holes. Of these, 
only those spacetimes which are asymptotically anti-de Sitter, i,e, 
have constant negative curvature at infinity, can be considered to be 
in stable thermal equilibrium, so long as their (outer) horizon 
radius exceeds $(-\Lambda)^{-1/2}$ where $\Lambda$ is the 
negative cosmological constant.\footnote{In fact we only consider 
very large black holes with $A \gg (-\Lambda)^{-1/2}$ for which 
(\ref{powa}) is assumed to be valid.} 

We may remark that there is no essential loss of generality in 
choosing the functional relation in eq.s (\ref{powa}, \ref{powp}) to 
be quadratic in the charge; in fact our analysis goes through for 
any positive power of Q.

Observe that the microcanonical ensemble definition of temperature is 
$T \equiv \left( S_{MC,E} \right)^{-1}$. On physical grounds one 
expects that $T \geq 0$ which implies a limit on the charge $Q^2 \leq 
(a \alpha/b \beta)~A^{\alpha + \beta}$ using eq. (\ref{powa}). This 
inequality is saturated by the extremal charge $Q_{ext}^2 = 
(a\alpha/b \beta) A^{\alpha + \beta}$ for which $T=0$. We also 
define, for future use, the ratio $q \equiv Q_0/Q_{ext}$ which must 
satisfy $-1 \leq q \leq 1$. 

The computation of the derivatives in eq. 
(\ref{delt2}) is now straightforward,
\begin{eqnarray}
A_{,E}|_{M,Q_0}~ &=& ~{\frac1a}~W(q, \alpha, \beta)~A^{1-\alpha} ~,~
A_{,EE}|_{M.Q_0}~=~ -\frac1{a^2}~X(q, \alpha, \beta)~A^{1-2\alpha} 
\nonumber \\
A_{,QQ}|_{M,Q_0} ~ &=&~ -\frac{b}{a}~Y(q, \alpha, 
\beta)~A^{1-\alpha-\beta} 
\nonumber \\
A_{,EQ}|_{M,Q_0}~ &=&~ \left( {b \over a^3} \right)^{1/2}~q~Z(q, 
\alpha, \beta)~A^{1-\frac32 \alpha -\frac12 \beta} ~.\label{partd}
\end{eqnarray}
The dimensionless functions $W$, $X$, $Y$ and $Z$ are explicitly 
calculable and are 
regular and non-vanishing for $0 \leq |q| < 1$; however they have 
poles at $q=\pm 1$ with $W \sim (1-q^2)^{-1}$ and $X~,~Y~,Z \sim 
(1-q^2)^{-3}$ as $|q| \rightarrow 1$. 

In terms of these functions, the leading correction is given by
\begin{eqnarray}
\Delta~=~\frac{C}{16}~\frac{b}{a}~{X~Y~-~q^2~Z^2 \over W^2}~ 
A^{-(\alpha + \beta)} ~,
\label{delt3}
\end{eqnarray}
so that the grand canonical entropy 
\begin{eqnarray}
S_G~=~{\tilde S}_{MC}~+~\frac12~(\alpha + \beta)~\log A~+~O(A^0) 
~\mbox{for $A >> 
(-\Lambda)^{-1}$ and $0 \leq |q| < 1$}~. \label{sg2}
\end{eqnarray} 
Note that the leading logarithmic correction is independent of $q$. 
However, as we approach extremality $|q| \rightarrow 1$, $\Delta$ 
diverges as 
$(1-q^2)^{-4}$, the $O(A^0)$ terms become large and the approximation 
of large horizon areas becomes invalid. This happens because as we 
approach extremality, $T \rightarrow 0$ and all fluctuations 
basically freeze out. Furthermore, as we shall argue below, the 
result also cannot be trusted for $q \rightarrow 0$, although for 
somewhat different reasons. 

\section{Relative fluctuations for large black holes}

Using the area law, we find, for adS black holes with $A >> \ell^2$,
\begin{eqnarray}
\Delta E^2 ~ &=& ~ - \frac14 ~\left( {A_{,QQ} \over \det ~ \Omega}
\right)_{M,Q_0}
~ \nonumber \\
\Delta Q^2 ~ &=& ~ -\frac14~ \left( {A_{,EE} \over \det~ \Omega }
\right)_{M,Q_0} ~,
\label{flu}   
\end{eqnarray}
where, eq. (\ref{deto}) is to be used for $\det \Omega$. 
Substituting the derivatives in eq. (\ref{partd}), we obtain
the rms fluctuations in mass and charge
\begin{eqnarray}
\Delta E~ &=& ~ 2a~\left( {Y \over X Y~-~q^2 Z^2}
\right)^{1/2}_{M,Q_0} A^{\alpha - \frac12}  \nonumber \\
\Delta Q ~ &=& ~ 2 \sqrt{{a \over b}}~ \left( {X \over X Y ~-~q^2
Z^2} \right)^{1/2}_{M,Q_0} A^{(\alpha + \beta -1)/2} ~.
\label{rms}
\end{eqnarray}
On the other hand, we have for the mean (equilibrium) values
\begin{eqnarray}
M~ = ~ a~ A^{\alpha} \left( 1~+~{\alpha \over \beta} q^2\right)
~\label{aq}
\end{eqnarray}
and
\begin{eqnarray}
Q_0~ = ~ q~\left( {a \alpha \over b \beta} \right)^{1/2}  A^{(\alpha
+ \beta)/2} ~. \label{eq}
\end{eqnarray}
Hence the relative fluctuations are
\begin{eqnarray}
{\Delta E \over M} ~ &=& ~ \left( 1~+~{\alpha \over \beta}~q^2
\right)^{-1} \left( {Y \over XY~-~q^2 Z^2}
\right)^{1/2} S_{BH}^{-1/2} ~ \nonumber \\
{\Delta Q \over Q_0} ~ &=& ~{1 \over q}~\left[ {\beta X \over \alpha
(X Y ~-~q^2 Z^2)} \right]^{1/2} S_{BH}^{-1/2} ~. \label{relf}
\end{eqnarray}
Thus, although the thermal fluctuation correction to the entropy has
a degree of non-universality, the relative fluctuations nevertheless
turn out fall off with $A$ for large $A$ in a universal manner,
scaling like $A^{-1/2}$ independent of the details of the energy-area
relation. This universal fall-off shows that the small fluctuation 
approximation employed in our analysis is valid for large area black 
holes as long as $q$ is not small. This requires the black hole 
charge to grow as a moderate proper fraction of the extremal charge 
which increases with area. 

However, for $|q| \rightarrow 0$, while the relative rms energy
fluctuation is bounded, the relative charge fluctuation {\it
blows up} as $q^{-1}$, thereby invalidating the small fluctuation
(Gaussian) approximation for charge. This happens not only for the
$Q_0 \sim 0$ case considered in ref. \cite{gm} (where it also follows
from the relation for the rms fluctuation in the charge quantum 
number, but appears to
have been overlooked) but for any value of $Q_0$ which does
not scale as $Q_{ext}(A)$ for large $A$.

\section{Discussion}

The primary result of this paper is the leading thermal fluctuation
correction $\frac12 (\alpha +\beta) \log A$ in the grand canonical
entropy (\ref{sg2}). It is obvious that when charge fluctuations are
taken into account in addition to energy fluctuations of a black
hole, the logarithmic correction due to such fluctuations depends on
the exponents appearing in the relation linking horizon area, mass
and charge. This is quite unlike the case studied in Ref.  
\cite{cm1} where the charge fluctuations were ignored; the energy
fluctuations in that case produced a correction $\frac12 \log A$ {\it
independent of the assumed area-energy relation}. In this sense,
charge fluctuations endow the thermal fluctuation correction with a
certain degree of `non-universality' absent in the earlier case.  
However, what is perhaps noteworthy is that the coefficient appearing
in the leading thermal correction to the area law is proportional to
the {\it sum of the exponents} appearing in the area-charge-energy
relation (\ref{powa}), and not related to any one of the exponents
alone. Another way of saying the same thing is that, if we use eq.  
(\ref{powp}) to express the area-charge-energy relation, the leading
correction is in fact independent of $\alpha$ and simply depends on
$\beta'$, which of course is the same as $\alpha + \beta$ . In
(\ref{powp}), the contribution of the charge is expressed as an
additional factor multiplying the energy-area relation used in the
canonical ensemble case \cite{cm1}. The fact that the leading thermal 
correction is proportional only to the exponent appearing in this 
additional factor reaffirms our earlier result of universality in 
absence of charge fluctuations. To this extent then, one could say 
that the leading thermal correction with charge fluctuations is at 
least `partly universal'. 

For the adS Reissner Nordstrom black hole, the area-charge-energy 
relation is given by
\begin{eqnarray}
E(A,Q) ~=~{1 \over 2 \ell^2}~\left( {A \over 4\pi} 
\right)^{3/2}~\left[ 
1~+~{4\pi \ell^2 \over A}~+~{(4\pi)^2 \ell^2 Q^2 \over A^2} \right] 
~, \label{ads}
\end{eqnarray}
where, $\ell \equiv (-\Lambda/3)^{-1/2}$. For $A \gg \ell^2$ and
moderate values of $|q|$ in the range $0 < |q| < 1$ ), we have here 
$\alpha=3/2~,~\beta'=2$, in the notation 
used in (\ref{powp}, implying that
the leading thermal correction is $\log A$. This result agrees with
that found in Ref. \cite{gm} for this black hole {\it in the limit}
$Q_0 \rightarrow 0$ (or equivalently $q \rightarrow 0$).  This limit
is however pathological since, as argued above, the mean-squared
charge fluctuation relative to its mean (equilibrium) value diverges
in this limit, thereby invalidating the Gaussian approximation which
has been used to compute the thermal fluctuation correction. We have
further demonstrated above that it is not at all necessary to go to
this small charge limit in order to compute the leading thermal
fluctuation correction. We may reiterate that our result eq. 
(\ref{sg2}) actually encompasses
a larger class of adS black holes which may include non-rotating adS
dilatonic black holes as well \cite{wilt}.

It follows that the mechanism proposed in Ref. \cite{med} to argue a
possible complete cancellation of logarithmic corrections in the
grand canonical computation of the entropy stands on loose ground. In
contrast, such a cancellation has already been shown earlier
\cite{cm2} between the logarithmic contribution in the $U(1)$
calculation of the microcanonical entropy \cite{km2} of non-rotating
black holes and the universal {\it canonical} correction found by us
earlier \cite{cm1}, both of which appear to be robust. Generalization
of these considerations to black holes with rotation stands
prominently on our agenda for the near future.

\end{document}